\documentclass[pss]{wiley2sp} 
\usepackage{amsmath}
\usepackage{amssymb}
\usepackage{epsfig}
\usepackage{bm}
\usepackage{pstricks}
\usepackage{pst-3d}

\newcommand{\tr}{\text{Tr}}

\newcommand{\be}{\begin{equation}}
\newcommand{\ee}{\end{equation}}

\begin{document}

%
%

\title{Entanglement of Hard-Core Bosons in Bipartite Lattices}

\author{Xue-Feng Zhang,\textsuperscript{\Ast,\textsf{\bfseries 1}} 
Raoul Dillenschneider,\textsuperscript{\textsf{\bfseries 2}} and 
Sebastian Eggert\textsuperscript{\textsf{\bfseries 2}}}

\institute{\textsuperscript{1}\,Department of Physics, Chongqing University, Chongqing 401331, People's Republic of China\\
\textsuperscript{2}\,Physics Dept.~and Res.~Center
OPTIMAS, Technische Universit\"at Kaiserslautern, 67663 Kaiserslautern, Germany}

\mail{e-mail
  \textsf{zxf198396@hotmail.com}}

%
%

\abstract{\bf The entanglement of hard-core bosons in square and honeycomb lattices with nearest-neighbor interactions
is estimated by means of quantum Monte Carlo simulations and spin-wave analysis. 
The particular $U(1)$-invariant form of the concurrence is used to establish a connection 
with observables such as density and superfluid density. For specific regimes the concurrence 
is expressed as a combination of boson density and superfluid density.}

\maketitle


\section{Introduction \label{Section1}}

The past few years have seen a large explosion of interest in the studies
of the interfaces between quantum information and many body systems.
Among the subjects of interest can be cited quantum information processing
in ultracold atomic gases \cite{Lewenstein}. This subject was initiated
by the first proposal of using ultracold atoms on optical lattices for
quantum information \cite{Jaksch}. The physics of quantum ultracold gases
in optical lattices has rapidly grown in interest \cite{Bloch}.
Recent theoretical developments suggest that ultracold gases may be used
for the experimental realization of the phenomenon of a supersolid 
\cite{Melko,Santos,Wessel,Becker,Eggert,rydberg,triang-ss,ss-imp}.

In another register, entanglement is an important element in quantum information. 
It is used in quantum computation \cite{NielsenChuang} and is also
a valuable resource in quantum thermodynamics \cite{EREuroPhys,Nature}.
Meanwhile it can characterize quantum phase transitions (QPTs) 
\cite{Sachdev,Osterloh,AmicoVedral,CaiZhouGuo,GuTianLin1,GuTianLin2,OsborneNielsen,QPT_RD,xxz}.
QPTs occur when the ground state of a many-body system 
at absolute zero temperature undergoes a qualitative change by variation of
a coupling and/or an external parameter \cite{Sachdev}.
The detailed analysis of QPTs is a very rich field and has the potential to 
uncover interesting physics and unexpected phase diagrams in a
variety of many-body quantum systems, such as models with spins, bosons, or fermions
on frustrated lattices \cite{triang-ss,xxz,deconfined,kagome-sf,tri-hubbard,j1j2}, 
quasi one-dimensional (1D) ladders \cite{ladder,honeycomb} 
and chains \cite{Osterloh,AmicoVedral,CaiZhouGuo,1d}, 
and unfrustrated geometries in 2D or higher 
dimensions \cite{rydberg,GuTianLin1,GuTianLin2,bkt,2d,swave}.
In this paper we will take a closer look at the signatures of 
QPT's in the pairwise entanglement between two sites 
for the example of hard-core bosons on
the square lattice and the honeycomb lattice.
Entanglement is maximal close to the critical points and its derivatives can 
signal more precisely the presence of a quantum phase transition at the critical 
points
\cite{Osterloh,AmicoVedral,CaiZhouGuo,GuTianLin1,GuTianLin2,OsborneNielsen,QPT_RD,xxz}.

Among the various quantities that can extract information on entanglement
can be mentioned the concurrence \cite{Wootters,OConnorWootters}, 
the quantum discord \cite{OllivierZurek,Zurek}, the entropy of entanglement 
\cite{Horodecki} and the negativity \cite{VidalWerner} for the most renowned.
Concurrence and quantum discord are pairwise measures of entanglement
while entropy of entanglement and negativity are bipartite measures.
Bipartite measures of entanglement are by construction \emph{isotropic} measures.
Pairwise measures of entanglement can be \emph{anisotropic} for the same system 
under considerations. The last point strongly motivates us to use concurrence to 
describe entanglement in hard-core bosonic models. Here we focus our attention 
to entanglement between hard-core bosons on square and honeycomb lattices
with nearest-neighbor interactions.

For particular cases the hard-core boson model with nearest-neighbor interactions 
in two spatial dimensions can be mapped onto a two dimensional spin-$1/2$ XXZ model \cite{Dorneich}. 
By means of quantum Monte Carlo simulations (QMC) and spin-wave analysis (SW), 
we estimate the entanglement by using concurrence. Concurrence appears to be the best choice to
study entanglement because it leads to pairwise entanglement and is easy to implement.
For a specific region of the phase diagram of the XXZ model the concurrence takes a very simple 
$U(1)$-invariant form \cite{Syljuasen}. This particular $U(1)$-invariant form 
is used to establish a connection with observables such as boson density and 
superfluid density of the hard-core boson system. Concurrence can henceforth
be expressed as a combination of boson density and superfluid density.

The outline of the paper is as follows.
In section \ref{Section2} we present the hard-core boson model and its mapping onto
the XXZ spin model.
In section \ref{Section3} we recall the elements of information 
theory which leads to the $U(1)$-invariant form of the concurrence.
In section \ref{Section4} the quantum Monte Carlo and spin-wave approaches
used to derive the quantum correlations are presented. 
In section \ref{Section5} we provide the results of QMC simulations and SW analysis.
In section \ref{Section7} the connection between concurrence and observables is established.
In section \ref{Section8} we conclude and discuss on potential outlooks.

\section{The model \label{Section2}}

We will consider a two dimensional system of hardcore bosons on the square lattice and
the honeycomb lattice. The Hamiltonian is given by
\begin{equation}
H = -t \sum_{\langle ij \rangle}(a^{\dagger}_i a_j + a^{\dagger}_j a_i) 
- \mu \sum_i {\hat n_i}
+ V \sum _{\langle ij \rangle} {\hat n_i} {\hat n_j},
\label{EqHam1}
\end{equation}
\noindent
where $\langle ij \rangle$ denotes nearest neighbor bonds, $a_i$
($a^{\dagger}_i$) destroys (creates) a hard-core boson on site $i$, and $\mu$ is
the chemical potential. The hopping parameter is denoted by $t$ and
the interaction between nearest neighbors is introduced by $V$.

To enforce the hard-core constraint in a simple way, the Hamiltonian \eqref{EqHam1}
is mapped onto the two dimensional XXZ model with external magnetic field.
The exact mapping is performed by $a_i^{\dagger} \leftrightarrow S_i^{+}$, 
$a_i \leftrightarrow S_i^{-}$, and ${\hat n_i}\leftrightarrow
S_i^z+1/2 \quad$ \cite{Dorneich,schmidt}. 
For the particular case with $V/2t = \Delta$ and $\mu = \lambda V$, where
$\lambda$ is half the coordination number $z$ ($z=3$ for a honeycomb lattice 
and $z=4$ for a square lattice), and in units of
$t/2$ the Hamiltonian reduces to the familiar XXZ spin model
\begin{eqnarray}
H_{XXZ} &=& \sum_{\langle i,j\rangle} \left[ -\left( \sigma^x_i \sigma^x_j + \sigma^y_i \sigma^y_j\right)
+ \Delta \sigma^z_i \sigma^z_j\right] + \kappa_\Delta,
\label{EqXXZ1}
\end{eqnarray}
\noindent
where the sum is taken over all nearest neighbor sites on a lattice which is bipartite. 
The operators $\sigma^\alpha$ with $\alpha = {x,y,z}$ are Pauli matrices and $\sigma^0$
is the unit matrix.
The constant $\kappa_\Delta$ is equal to $-\Delta z N /8$ and simply arises from 
the mapping between the spins and bosons operators. $N$ is the number of sublattice
spins of the bipartite honeycomb lattice.
We keep explicitly the constant $\kappa_\Delta$
present in the Hamiltonian \eqref{EqXXZ1} because it will be important for the computation
of spin-spin correlations functions.
The Hamiltonian is real and invariant under $U(1)$ rotation about the spin $z$ axis.
This continuous symmetry can only be spontaneously broken in dimensions higher 
than one and for $|\Delta| < 1$. A global $Z_2$ symmetry about the spin $x$ (or $y$) 
is also present.

At the critical point $\Delta_c = 1$ the XXZ spin lattice undergoes
a quantum phase transition between an XY phase for $-1 < \Delta < 1$ 
and an Ising antiferromagnetic phase for $\Delta > 1$.
For $\Delta < -1$ the XXZ system is in a ferromagnetic phase.

In order to extract information about entanglement in the system by means of 
concurrence, we need to build the joint state of two spin sites. 
The two-site density matrix provides such requirement.

\section{Concurrence \label{Section3}}

The information on the joint state is
contained in the two-site density matrix $\rho_{ij}$ which
is derived from the following operator expansion \cite{OsborneNielsen}
\begin{eqnarray}
\rho_{ij} = \tr_{\bar{ij}} \left[ \rho \right] 
= \frac{1}{4} \sum_{\alpha,\beta=0}^3
\Theta_{\alpha \beta} \sigma_i^\alpha \otimes \sigma_j^\beta,
\end{eqnarray}

\noindent
where the trace is taken over the whole system excluding the sites
$i$ and $j$. The coefficients $\Theta_{\alpha \beta}$ of the expansion are 
related to the spin-spin correlation functions through the relation
\begin{eqnarray}
\Theta_{\alpha \beta} 
= \tr \left[\sigma_i^\alpha \sigma_j^\beta \rho_{ij} \right]
= \langle \sigma_i^\alpha \sigma_j^\beta  \rangle.
\end{eqnarray}

\noindent
Owing to the symmetry of the Hamiltonian most of the coefficients 
$\Theta_{\alpha \beta}$ are equal to zero. Translation invariance requires that
the density matrix $\rho_{ij}$ is only a function of distance $r=i-j$  
independent of the position $i$. 
The reflection symmetry leads to $\rho_{ij} = \rho_{ji}$,
the Hamiltonian being real the density matrix verifies $\rho_{ij}^{*} =
\rho_{ij}$. Combining all symmetry
constraints the density matrix expressed in the natural
basis $\left\{|\downarrow \downarrow \rangle, |\downarrow \uparrow\rangle, |\uparrow \downarrow \rangle
, |\uparrow \uparrow\rangle \right\}$ reduces to
\begin{eqnarray}
\rho_{ij}
= \left(
\begin{array}{cccc}
u & g & g & y \\
g & w & x & g \\
g & x & w & g \\
y & g & g & u
\end{array}
\right),
\label{EqReducedRho}
\end{eqnarray}

\noindent
where the matrix elements are given by
$
u = \frac{1}{4} + \frac{\langle \sigma^z_i \sigma_j^z\rangle}{4}
$,
$w = \frac{1-\langle \sigma^z_i \sigma^z_j \rangle}{4}$,
$x = \frac{\langle \sigma_i^x \sigma^x_j\rangle 
+ \langle \sigma_i^y \sigma^y_j\rangle}{4}$,
$g = \frac{\langle \sigma^x\rangle}{4}$,
and
$y = \frac{\langle \sigma_i^x \sigma^x_j\rangle 
- \langle \sigma_i^y \sigma^y_j\rangle}{4}$.  
Therefore, information on entanglement of the system can be extracted easily 
from the two-point correlators.

As mentioned before, a good indicator of entanglement is provided 
by the concurrence $\mathcal{C}$.
The concurrence of two spins may be computed from the joint state 
$\rho_{ij}$ through the formula $\mathcal{C} = \max \left\{ 0, \gamma_1 - \gamma_2 
-\gamma_3 -\gamma_4 \right\}$ where the $\gamma_i$ are the eigenvalues
in decreasing order of the matrix $R = \sqrt{\rho_{ij} \widetilde{\rho_{ij}}}$
 \cite{Wootters,OConnorWootters}.
The square root of the eigenvalues of the matrix $R$ are given by
\begin{eqnarray}
\Gamma_{\pm}\!  &=& \! \frac{1}{4} \left| \sqrt{\left(1 + \langle \sigma^x_i \sigma^x_j\rangle \right)^2 \! -\! 4 \langle \sigma^x \rangle}
\pm |\langle \sigma^y_i \sigma^y_j\rangle\! -\! \langle \sigma^z_i \sigma^z_j\rangle| \right| \notag
 \\
\Theta_{\pm}\! &=&\! \frac{1}{4} \left| 1 - \langle \sigma^x_i \sigma^x_j\rangle 
\pm \left( \langle \sigma^y_i \sigma^y_j\rangle + \langle \sigma^z_i \sigma^z_j\rangle  \right) \right|.
\end{eqnarray}
For two-space dimensions and for $|\Delta| < 1$ the average $\langle \sigma^x \rangle$ takes spontaneous
non-zero values, the $U(1)$ symmetry of the XXZ model is spontaneously broken.
According to Sylju{\aa}sen \cite{Syljuasen} the concurrence in the symmetry-broken state \eqref{EqReducedRho}
can take the invariant $U(1)$ form 
\be
\mathcal{C} = \frac{1}{2} \left( \langle \sigma^x_i \sigma^x_j\rangle 
+ \langle \sigma^y_i \sigma^y_j\rangle -\langle \sigma^z_i \sigma^z_j\rangle -1  \right)
\ee
if and only if the spin-spin correlation functions verify
$\langle \sigma^y_i \sigma^y_j\rangle + \langle \sigma^z_i \sigma^z_j\rangle
> \langle \sigma^x_i \sigma^x_j\rangle - 1$ and 
$\langle \sigma^y_i \sigma^y_j\rangle > \langle \sigma^z_i \sigma^z_j\rangle$.

The $U(1)$-invariant form of the concurrence is of particular importance for two reasons.
The first reason is that it allows computation of concurrence by means of quantum Monte Carlo simulations.
The second reason is that it helps to establish a connection between entanglement and observables
such as boson density and superfluid density. As will be shown later
the concurrence can be expressed as a linear combination of boson density and superfluid density.

To work out the correlation functions we will apply both an 
analytical approach using spin-wave theory and a numerical approach 
with quantum Monte Carlo methods.

\section{Spin-wave analysis and Quantum Monte Carlo simulations \label{Section4}}

Spin-wave analysis provides a good analytical approach as was shown
for the hardcore bosons problem on the
square lattice \cite{Dorneich} and the effective 
honeycomb lattice \cite{Eggert}. A spin-wave analysis can be performed in the
regime $-1 < \Delta < 1$ for which the ground state corresponds to spins aligned 
in any direction within the XY plane. In order to grab fully the particular 
symmetry of the XY interactions a Haldane mapping is performed \cite{Auerbach} and a 
semi-classical approach can be applied \cite{Pires} to compute the
spin-spin correlation function.

We first recall the transformations applied to the Hamiltonian \eqref{EqXXZ1} 
that lead to a diagonalized Hamiltonian as demonstrated in Ref. \cite{Pires}.
The following demonstration is in some steps very similar to the derivation of 
the non-linear sigma model \cite{Auerbach}. We are going to work out the spin-spin 
correlations function by means of a semi-classical version of the 
Hamiltonian \eqref{EqXXZ1}.

First the spin operators are expressed by means of the Haldane mapping.
In terms of the in-plane angular coordinate $\phi_i$ and the spin projection 
$\sigma^z_i$ in the direction $Oz$, the spin operators read
\begin{eqnarray}
\vec{\sigma}_i = \left(\sqrt{1-{\sigma^z_i}^2} \cos{\phi_i}, 
\sqrt{1-{\sigma^z_i}^2} \sin{\phi_i},
\sigma^z_i\right).
\end{eqnarray}

\noindent
With this mapping the Hamiltonian \eqref{EqXXZ1} becomes
\begin{eqnarray}
H = \sum_{\langle i,j\rangle} \Big(
&-&\sqrt{\left(1 - {\sigma^z_i}^2 \right)
\left(1 - {\sigma^z_j}^2 \right)} \cos{ \left(\phi_i - \phi_j \right)}
\notag \\
& +& \Delta \sigma^z_i \sigma^z_j \Big) + \kappa_\Delta,
\label{EqHamHaldane}
\end{eqnarray}

\noindent
At zero temperature we can assume a dilute spin-wave boson gas.
In this case, the Hamiltonian can reasonably be expanded for small
$\sigma^z$ and $\phi$. The expansion 
of the Hamiltonian 
up to quadratic terms and to second order in $\sigma^z$ and $\phi$
reads
\begin{eqnarray}
H^{(2)} = \sum_{\langle i,j\rangle} \Big(
&-&1 + \frac{1}{2} \left({\sigma^z_i}^2 + {\sigma^z_j}^2\right)
+ \frac{1}{2} \left(\phi_i - \phi_j\right)^2
\notag \\
& +& \Delta \sigma^z_i \sigma^z_j \Big) + \kappa_\Delta.
\label{expansion}
\end{eqnarray}

\noindent
Higher orders of the expansion are not explicitly considered. Later in
the derivation of the spin-spin correlation functions we introduce corrections
arising from those neglected higher order terms. After Fourier transformation 
the Hamiltonian $H^{(2)}$ becomes
\begin{eqnarray}
H^{(2)} &=& \sum_k \Big(
\left( 1 - |\gamma_k| \cos{\varphi_k}\right) \phi_k \phi_{-k}
\notag \\
&&
\ \ +
\left( 1 + \Delta |\gamma_k| \cos{\varphi_k}\right) \sigma^z_k \sigma^z_{-k}
 \Big) + \kappa_\Delta + \kappa_0,
\end{eqnarray}

\noindent
where $\kappa_0 = -zN$ collects the constant parts of $H^{(2)}$.
We also introduced the structure factor $\gamma_k = \frac{1}{z} \sum_d e^{i \vec{k}.\vec{r}_d} 
= |\gamma_k| e^{i \varphi_k}$ which is a complex number for the honeycomb lattice 
and where the sum runs over 
nearest neighbors sites. The amplitude of the structure factor  for the
honeycomb lattice reads 
\be 
|\gamma_k| = \frac{1}{3} \left[3 + 4 \cos \frac{3 k_x}{ 2}  \cos \frac{\sqrt{3} k_y}{2} 
+ 2 \cos \left(\sqrt{3} k_y \right)\right]^{1/2}.\ee
For square lattices the phase $\varphi_k$
equals zero and the structure factor is 
given by $\gamma_k = \left( \cos{k_x} + \cos{k_y} \right)/2$. 
We then use the canonical transformation
$\phi_k = \alpha_k \left(b^\dagger_k + b_{-k} \right)$ and
$\sigma^z_k = i \beta_k \left(b^\dagger_k - b_{-k} \right)$ where
$b_k$'s are bosons and
\begin{eqnarray}
\alpha_k & = & \left[
\frac{1+\Delta |\gamma_k| \cos{\varphi_k}}{1-|\gamma_k| \cos{\varphi_k}}
\right]^{1/4},\notag \\
\beta_k & =&  
\left[
\frac{1-|\gamma_k| \cos{\varphi_k}}{1+\Delta |\gamma_k| \cos{\varphi_k}}
\right]^{1/4}.
\end{eqnarray}
Finally, the Hamiltonian takes the diagonalized form
\begin{eqnarray}
H^{(2)} = 
\kappa_0 + \kappa_\Delta +
\sum_k \omega_k \left( b^\dagger_k b_k + 1/2\right),
\label{EqHam2}
\end{eqnarray}

\noindent
where $\omega_k = 4 z \left[\left( 1 - |\gamma_k| \cos{\varphi_k}\right) 
\left(1 + \Delta |\gamma_k| \cos{\varphi_k}\right) \right]^{1/2}$.

In order to compute the spin-spin correlations functions we can employ 
the Hellmann-Feynman theorem which relates the correlations functions
to the bond-energy of the system \cite{Hellman,Feynman}. The bond-energy $e(\Delta)$ is 
defined as the average of the Hamiltonian $H$ divided by the number of bonds, 
$e(\Delta) = \langle H \rangle/zN$. The spin-spin correlations functions
are easily given by $\langle \sigma^z_i \sigma^z_j\rangle =
\frac{\partial e(\Delta)}{\partial \Delta}$ and 
$\langle \sigma^x_i \sigma^x_j + \sigma^y_i \sigma^y_j \rangle 
= -e(\Delta) + \Delta \frac{\partial e(\Delta)}{\partial \Delta}$.
Taking the average of the second order Hamiltonian $H^{(2)}$ over the ground state
leads to the approximated ground state energy $e^{(2)} = \langle H^{(2)} \rangle/zN$.
The spin-spin correlations functions can then be expressed in terms of the
approximated bond energies 
\begin{eqnarray}
\langle \sigma^z_i \sigma^z_j\rangle &\simeq&
\frac{\partial e^{(2)}(\Delta)}{\partial \Delta} 
+ \kappa_{zz},
\label{EqSzSz2} 
 \\
\langle \sigma^x_i \sigma^x_j + \sigma^y_i \sigma^y_j \rangle 
&\simeq& 
-e^{(2)}(\Delta) + \Delta \frac{\partial e^{(2)}(\Delta)}{\partial \Delta}
+ \kappa_{xy}.
\label{EqSpSm2}
\end{eqnarray}

\noindent
Non-negligible corrections to the spin-spin correlations functions 
\eqref{EqSzSz2} and \eqref{EqSpSm2} may be taken into account from
higher order terms of the expansion of the Hamiltonian \eqref{EqHamHaldane}.
These corrections, $\kappa_{zz}$ and $\kappa_{xy}$, are functions
of the anisotropic parameter $\Delta$. In the region $-1< \Delta < 1$, 
we approximate $\kappa_{zz}$ and $\kappa_{xy}$ by constants. 
For the honeycomb lattice
 we find $\kappa_{zz} \simeq 0.15$ and $\kappa_{xy} \simeq 1.65$, while
for the square lattice we obtain $\kappa_{zz} \simeq 0.5$ and $\kappa_{xy} \simeq 2$
from fits to 
the Monte Carlo simulation results at
$\Delta = -1$. Despite the aggressive approximations we have applied here 
we will see that a reasonable description of the system is obtained.

The QMC simulations used in the present work are based 
on the stochastic series expansion algorithm \cite{Sandvik1,Sandvik2,Sandvik3}. 
The numerical results are obtained for lattices of size $L \times L$ 
($L=12$, 18 and 24 for honeycomb lattices, and $L=16$, 20 and 24 for square lattices) 
with periodic boundary conditions and a finite inverse temperature of $\beta t = 50$.

\section{Results \label{Section5}}

\begin{figure}[t]
\epsfig{file=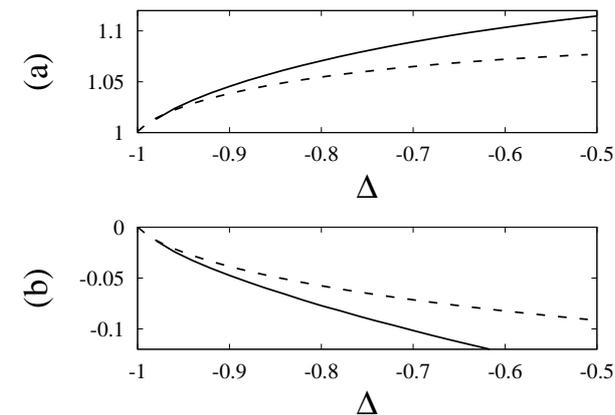,width=\columnwidth}
\caption[]{\label{Fig1Hexa} Spin-spin correlation functions for nearest neighbors on the honeycomb lattice,
with 
(a) $\langle \sigma^x_i \sigma^x_j + \sigma^y_i \sigma^y_j\rangle$ 
and (b) $\langle \sigma^z_i \sigma^z_j\rangle$ 
obtained from
Monte-Carlo simulation (full line) and spin-wave analysis (dashed line).}
\end{figure}

Figures \ref{Fig1Hexa}(a) and \ref{Fig1Hexa}(b) provide a comparison between the spin-spin correlation
functions derived from quantum Monte Carlo simulations and the
spin-wave approach on the honeycomb lattice.
For the region $\Delta < 0$ spin-wave theory provides a good approximation
of the correlation functions.   For larger  
anisotropic parameter $\Delta$ the deviations increase slowly, since the interaction between 
spin waves become more and more relevant.  
Note that the correlation function 
$\langle \sigma^x \sigma^x + \sigma^y \sigma^y\rangle$ is larger than one for 
$\Delta > -1$. This may surprise at a first glance however considering the two 
inequalities $-1 \le \langle \sigma^x \sigma^x \rangle \le 1$ and 
$-1 \le \langle \sigma^y \sigma^y \rangle \le 1$ we expect that 
$\langle \sigma^x \sigma^x + \sigma^y \sigma^y\rangle$ belongs to the range 
$\left[-2,2\right]$. 
In particular, for the spin singlet and triplet $|T_{\pm}\rangle = \left( 
| \uparrow \downarrow \rangle \pm | \downarrow \uparrow \rangle \right)/\sqrt{2}$ 
the XY-correlation function reads $\langle T_{\pm} | 
\sigma^x \sigma^x + \sigma^y \sigma^y| T_{\pm} \rangle = \pm 2$.
This supports the fact that the absolute value of the XY-correlation function 
can take values larger than one.

\begin{figure}
\epsfig{file=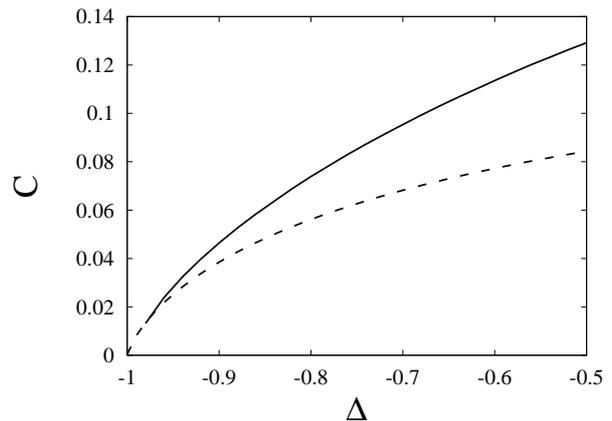,width=\columnwidth}
\caption{\label{Fig2Hexa} Concurrence derived from Monte-Carlo simulation (full line) 
and spin-wave analysis (dashed line) on the honeycomb lattice.}
\end{figure}

Figure \ref{Fig2Hexa} depicts the concurrence obtained from quantum Monte Carlo on honeycomb lattices.
For $\Delta < -1$ the spin systems is in a ferromagnetic phase. The state of the system can be
expressed as a product of separate states, the concurrence is equal to zero and the
system is separable. For $\Delta \ge 1$, the system is in an antiferromagnetic
phase. For $\Delta = 1$ the concurrence is maximal and the system is maximally entangled.
Increasing the parameter $\Delta$ from the ferromagnetic to the antiferromagnetic phase 
increases the entanglement of the spin system (or equivalently the hard-core boson system).

The concurrence 
obtained from spin-wave approach on honeycomb lattices
(dashed line in Figure \ref{Fig2Hexa}) 
agree with QMC predictions close to the ferromagnetic phase 
transition, for $\Delta \rightarrow -1$. For larger values of $\Delta$ the boson gas 
of spin-wave excitations becomes denser. Hence the approximation of a dilute gas no longer holds
and our present spin-wave is no longer valid.

\begin{figure}
\epsfig{file=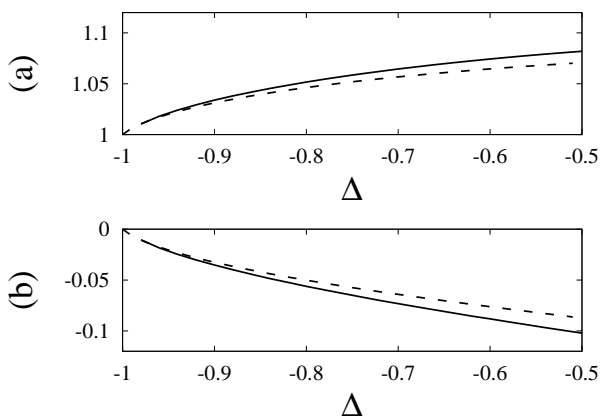,width=\columnwidth}
\caption{\label{Fig1Sq} Spin-Spin correlation functions for nearest neighbors on the square lattice,
(a) $\langle \sigma^x_i \sigma^x_j + \sigma^y_i \sigma^y_j\rangle$ 
and (b) $\langle \sigma^z_i \sigma^z_j\rangle$ 
obtained from
Monte-Carlo simulation (full line) and spin-wave analysis (dashed line).}
\end{figure}

\begin{figure}
\epsfig{file=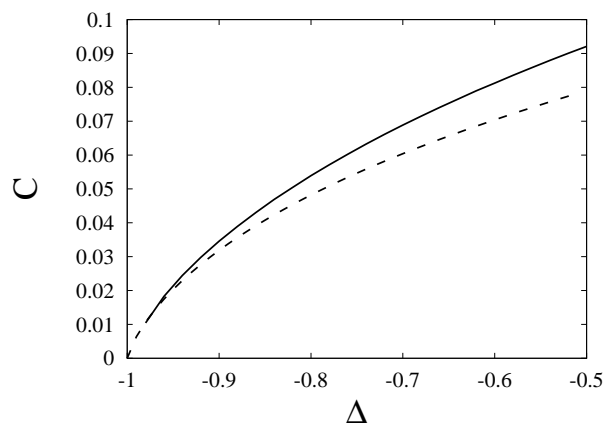,width=\columnwidth}
\caption{\label{Fig2Sq} Concurrence derived from Monte-Carlo simulation (full line) 
and spin-wave analysis (dashed line) on the square lattice.}
\end{figure}

Similar results are obtained for hard-core bosons on square lattices as depicted in figures
\ref{Fig1Sq} and \ref{Fig2Sq}.

\section{Concurrence, boson density and superfluidity \label{Section7}}

According to the mapping relating the hard-core bosons and spins,
the $U(1)$-invariant form of the concurrence can be expressed in terms 
of the boson density and superfluid density of the hard-core boson system. 

Indeed the spin-spin correlation function in the z direction can be easily expressed 
in terms of the boson density by making use of the exact mapping ${\hat n_i}\leftrightarrow
S_i^z+1/2$ which leads to $\langle S^z_i S^z_j \rangle 
= 1/4 - \langle \hat n \rangle + \langle \hat n_i \hat n_j\rangle$.

The superfluid density is related to the energy cost to introduce a twist $\nu$
between pairs of nearest neighbors spins. The superfluid density is given by the second derivative 
of the energy of the spin system with respect to the twist $\nu$, 
$\rho_s = d^2 \langle H(\nu) \rangle / d^2 \nu \quad$ \cite{Schulz,Cuccoli}.
The Hamiltonian $H(\nu)$ is derived from the XXZ model by application of a 
local rotation at site $i$ by an angle $\nu_i$ around the $z$ axis,
$S_i^{+} \rightarrow S_i^{+} e^{i\nu_i}$, $S_i^{-} \rightarrow S_i^{-} e^{-i\nu_i}$
and $S_i^z \rightarrow S_i^z$. Expanding the Hamiltonian around $\nu_{ij} = \nu_i -\nu_j = 0$
leads to $H(\nu) = H + \sum_{\langle i,j \rangle} \left(\nu_{ij} J^{s}_{ij}
+ \nu_{ij}^2/2 T_{ij} \right)$, where $J^{s}_{ij} = (i/2) t \left( S_i^{+}S_j^{-} - S_j^{+}S_i^{-} \right)$
is the spin current in the $z$ direction and $T_{ij} = (1/2) t \left( S_i^{+}S_j^{-} + S_j^{+}S_i^{-}\right)$
is the spin-kinetic energy \cite{Schulz,Cuccoli,Lecheminant}. In first-order perturbation theory the spin stiffness
is given by $\rho_s = \frac{1}{2N} \frac{\partial^2}{\partial \nu^2} \sum_{ij} \nu_{ij}^2 \langle T_{ij} \rangle$.
Second order perturbation leads to a term integrating the current-current correlator, with respect to $J_s$, that
is neglected in our spin-wave approach. The spin stiffness for a uniform twist $\nu$ and a given direction leads to
$\langle S_i^x S_j^x + S_j^y S_i^y \rangle \simeq 2 \rho_s /t$.

Replacing the spin-spin correlation functions by their linear expressions with respect
to the boson density and superfluid density the concurrence reads

\begin{eqnarray}
\mathcal{C} &\simeq& \max \left\{0, K_0 + K_b \rho_b + K_s \rho_s + K_{cor,b} \langle \rho_b \rho_b \rangle \right\},
\end{eqnarray}

\noindent
where $K_0 = -1$, $K_b = K_{cor,b} = 2$ and $K_s = 4/t$ are constants. 
This expression provides a direct way to approximately measure the entanglement between 
two hard-core bosons experimentally.

\section{Conclusion \label{Section8}}

In conclusion we proposed an analytical as well as numerical approach
to estimate the entanglement in a hard-core boson model on the honeycomb lattice and the square lattice. 
By means of the particular $U(1)$-invariant form that the concurrence takes we compare
spin-wave theory and quantum Monte Carlo measures of entanglement.
Moreover, we also show the existence of an approximate linear relation
between concurrence, boson density and superfluid density.
This relation may be used in experimental measurements for direct
evaluation of the entanglement in hard-core boson system.

It has to be mentioned that the present demonstration is generic and may be easily 
applied to different lattice symmetries. The $U(1)$-form of the concurrence is
sensitive to the lattice symmetry only through the spin-spin correlation functions.
The linear relation of the concurrence with observables should also remain valid.

\begin{acknowledgement}
This work was supported by the DFG via the Research
Center Transregio 49. R.D. would like to thank A. Metavitsiadis
for the helpful discussions on the present subject.
\end{acknowledgement}


\end{document}